\documentclass[aps,nofootinbib,groupaddress,floatfix,preprint]{revtex4-1}
\usepackage{graphicx}
\usepackage[left]{lineno}
\usepackage{amsmath,amsthm,amssymb,latexsym,amsfonts,times,mathrsfs,verbatim}
\usepackage{bm}
\usepackage{color}
\usepackage{gensymb}
\begin{document}

\setcitestyle{super}

\title{Magnetic Actuation and Feedback Cooling of a Cavity Optomechanical Torque Sensor} 

\author{P.H. Kim}
\author{B.D. Hauer}
\author{T.J. Clark}
\author{F. Fani Sani}
\author{M.R. Freeman}
\author{J.P. Davis}\email{jdavis@ualberta.ca}
\affiliation{Department of Physics, University of Alberta, Edmonton, Alberta, Canada T6G 2E9}

\begin{abstract}  
We demonstrate the integration of a mesoscopic ferromagnetic needle with a cavity optomechanical torsional resonator, and its use for quantitative determination of the needle's magnetic properties, as well as amplification and cooling of the resonator motion. With this system we measure torques as small as 32 zNm, corresponding to sensing an external magnetic field of 0.12 A/m (150 nT). Furthermore, we are able to extract the magnetization (1710 kA/m) of the magnetic sample, not known a priori, demonstrating this system's potential for studies of nanomagnetism. Finally, we show that we can magnetically drive the torsional resonator into regenerative oscillations, and dampen its mechanical mode temperature from room temperature to 11.6 K, without sacrificing torque sensitivity.
\end{abstract}  

\maketitle

{While the field of cavity optomechanics has enabled remarkable advances in measuring and manipulating nanomechanical resonators at, or approaching, their quantum limits,\cite{Teu11,Cha11} the most attractive applications of cavity optomechanics require integration with one or more additional systems.  For example, the first definitive test of quantum behaviour in a mechanical resonator was enabled by integration of a microwave optomechanical resonator with a superconducting qubit.\cite{Oco10}  Further experiments have demonstrated coupling between optomechanical resonators and atomic gases,\cite{Cam11} superfluids,\cite{Har16} and magnetic materials,\cite{Forstner12,Wu17} with the possibility of coupling to spins\cite{Rath13,Hoang16,Mitchell16,Burek16} and even biological samples.\cite{Li16} Such hybrid systems could be the key to wavelength conversion at the quantum level,\cite{Hil12} and enable ultra-sensitive measurements of nano- and mesoscale samples.\cite{Los15}  Here, we demonstrate integration of a mesoscopic ferromagnetic needle with a cavity optomechanical resonator, and show that this hybrid system can be used for magnetic field sensing, with a thermal limit of 0.12 A/m (150 nT); quantitative determination of the needle's magnetization; and magnetic feedback cooling of the mechanical mode from room temperature to below 12 K -- representing the first demonstration of feedback cooling with a magnetic cavity optomechanical system.  Future applications include high-frequency measurements of collective spin excitations in nanoscale materials\cite{Los15} and broadband microwave-to-telecom wavelength conversion.\cite{Bochmann13}}

\begin{figure}[b]
\centerline{\includegraphics[width=4.8in]{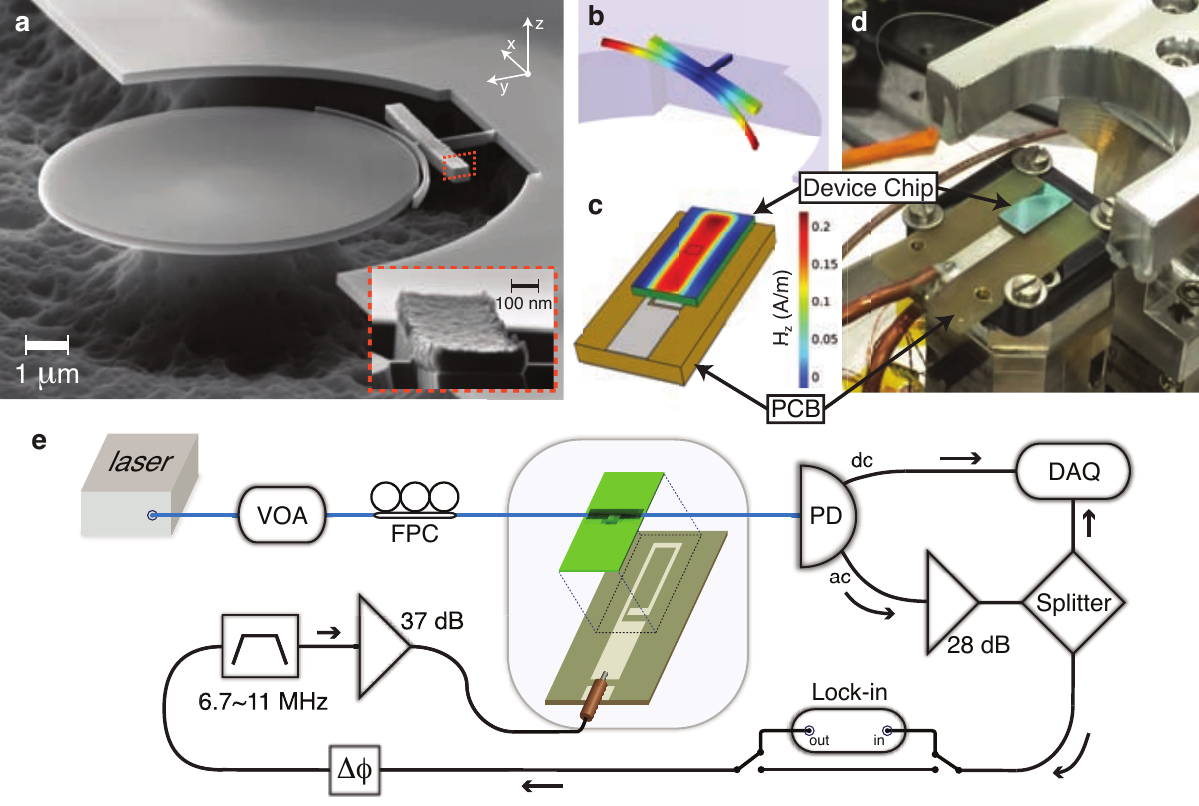}}
\caption{{\label{fig1}}\textbf{Magneto-optomechanical system.} \textbf{a}, Scanning electron microscope (SEM) image, tilted to 70 degrees, showing the optomechanical torque sensor, with an 8.8 $\mu$m diameter optical WGM microdisk evanescently coupled to an arced torsional resonator. Inset shows the integrated trilayer needle of Cr, Fe, and Cr, which has a total thickness of 83 nm (4.39 $\mu$m long and 410 nm wide), near the end of the torsion-arm.  \textbf{b}, A simulation of the torsional mode at $\Omega_{\rm m}/2\pi = 7.2$ MHz ($I_{\rm eff}$ = 4.4 pg$\cdot \mu$m$^2$, $m_\textrm{eff} = 445$ fg), which can be driven by a magnetically induced torque along the $y$-axis.  \textbf{c}, Simulation of the out-of-plane magnetic field along the $z$-axis at the top surface of the device chip, with 1 mA of applied current.  The device position is indicated by the black square, chosen for large field strength and relative field uniformity. \textbf{d}, Photograph and \textbf{e}, schematic of the optomechanical transduction and magnetic feedback.  (VOA = variable optical attenuator, PD = photodetector, DAQ = data acquisition system, FPC = fibre polarization controller).  The phase of the feedback signal is measured using the high-frequency lock-in amplifier and is varied by the length of coaxial cable. The shaded blue region depicts the contents inside the vacuum chamber (see Methods): the optomechanical chip, the printed-circuit-board (PCB) drive chip, and the dimpled tapered fibre.}
\end{figure}

One of the main considerations in the design of hybrid systems is taking advantage of the best properties of the individual systems, without sacrificing their performance during integration.  Hence we have chosen to directly integrate a magnetic structure onto our mechanical resonator.  The architecture of our magneto-optomechanical system is shown in Fig.\,1.  The mechanical structure, with a low effective moment of inertia torsional mode (see Fig.\,1b), is separated from the evanescent field of a whispering-gallery-mode (WGM) optical cavity by an 87 nm vacuum gap.  A platform for the magnetic sample was designed near the end of the torsion arm, amplifying the magnetic actuation of the torsional resonator,\cite{Kim16} yet is sufficiently far from the evanescent field of the WGM such that its optical properties are unaffected ($Q_\textrm{opt} = 5.3\times 10^4$).  On this platform we have deposited a ferromagnetic sample, which enables the mechanical motion to be driven, amplified, or dampened, by an alternating (ac) external magnetic field.  Details on device fabrication and measurement can be found in the Methods section.  

\begin{figure*}[t]
\centerline{\includegraphics[width=7.07in]{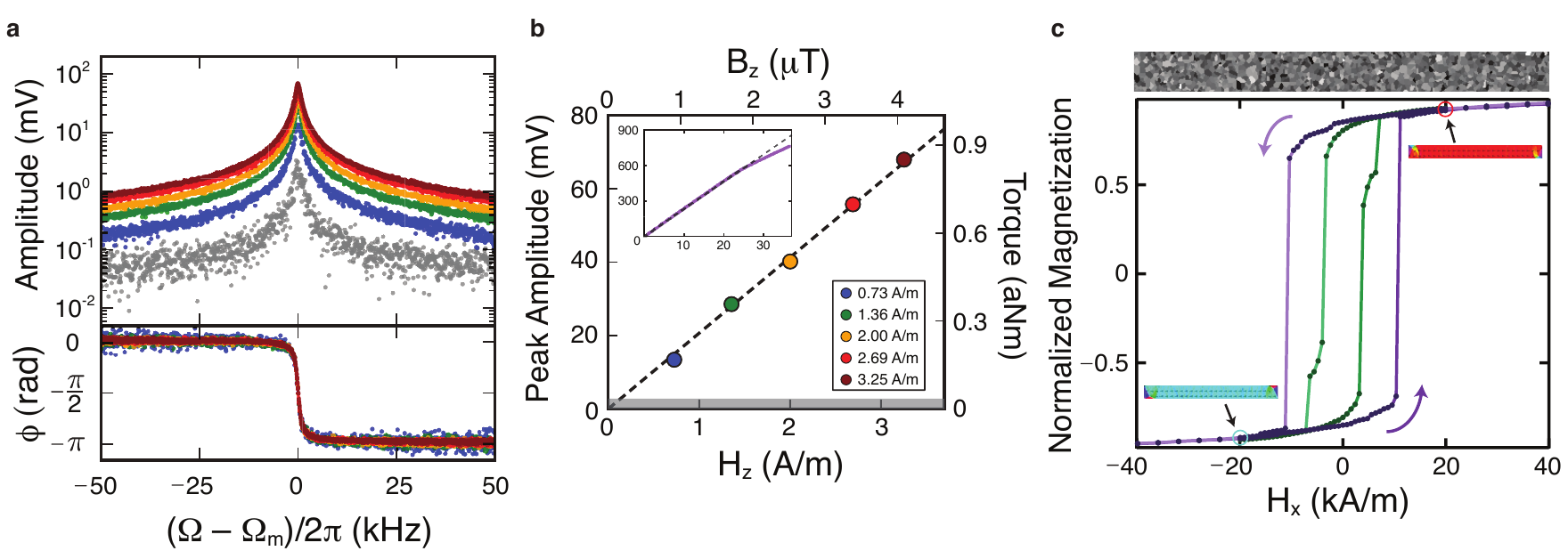}} 
\caption{{\label{fig2}} \textbf{Magnetic actuation and sensing.} \textbf{a}, Amplitude of the thermomechanical (grey) and magnetically actuated (coloured) torsional resonance, with corresponding phase of the driven traces.  Colours in \textbf{a} correspond to the drive magnetic field plotted in panel \textbf{b}, which shows the relationship between the peak mechanical signal amplitude (left axis) and calibrated torque (right axis), vs the magnetic drive.  The grey band represents the thermomechanically-limited minimum field sensitivity of 0.12 A/m, corresponding to a thermal torque of 32 zNm.  Inset, with the same axes, shows the deviation of the continuously-measured peak response (purple) from the linear fit extrapolated from low-field (dashed).  Above 25 A/m the optical resonance is shifted due to heating from the drive chip.  \textbf{c}, Simulated magnetization (normalized to $M_\textrm{s}$) hysteresis (see Methods for details) with in-plane domain structure at circled points.  At high fields the needle is nearly saturated, with triangular domains at each end.  As the field is lowered these domains move towards the centre, reducing the net magnetization. For a uniform iron film (green trace) at zero field, the remanent magnetization corresponds to 79\% of the saturation moment.  Adding polycrystalline grains, as shown above panel c, increases the remanent magnetization to 85\%, as seen in the purple trace.}
\end{figure*}

To test the responsiveness of the torsional resonator to magnetic actuation, we first characterized the driven response.  Since the ferromagnetic needle is magnetized, applying an external magnetic field, $\bm{H}$, perpendicular to its magnetic moment, $\bm{m}$, causes a torque, 

\begin{equation}
\bm{\tau} = \bm{m} \times \mu_0\bm{H}.
\label{torque}
\end{equation}

\noindent
Here, the magnetic moment is along the $x$-axis.  We apply an orthogonal ac magnetic field along the $z$-axis that generates a torque along the $y$-axis, {\it i.e.} the torsional axis of the resonator, Fig.\,1.  Sweeping the frequency of the magnetic drive through the mechanical resonance results in magnetically-actuated torsional motion, as shown in Fig.\,2.  The increase in peak amplitude of the mechanical spectrum is linear with magnetic drive until 25 A/m (31 $\mu$T), shown inset to Fig.\,2b.  This allows us to calibrate the peak amplitude in terms of the external magnetic field, and determine the thermomechanically-limited minimum field sensitivity of 0.12 A/m (150 nT).  Therefore as a magnetic field sensor, this system is linear over two orders of magnitude, from 0.12 A/m to 25 A/m, with a responsivity of 168 $\pm$ 2 $\mu$rad/(A/m). Note that lowering the temperature of the resonator would improve its dynamic range and sensitivity.\cite{Kim16}

Using the thermomechanical calibration method outlined in Ref.~\citenum{Hauer13}, we determine the measured Brownian torsional motion of the mechanical resonator, and hence the thermal torque applied to it. The right axis of Fig.\,2b shows the calibrated torque measured at the mechanical resonance frequency, with a minimum (thermal) torque of 32 zNm. In addition, the straight line fit to the data in Fig.\,2b using Eqn.\,\ref{torque}, allows extraction of the total magnetic moment of the ferromagnetic needle, $\bm{\left|m\right|} = (2.06 \pm 0.02)\times10^{-13}$ Am$^2$, or equivalently, $2.2 \times10^{10} \mu_\textrm{B}$.  Taking into account the measured volume of the iron portion of the needle, $V=(1.21 \pm 0.09) \times10^{-19}$ m$^3$, we find a magnetization, $M$, of 1710 $\pm$ 140 kA/m.  This compares favourably with the room temperature saturation magnetization of bulk iron, $M_{\rm s}^\textrm{iron}=1710$ kA/m.\cite{Wohlfarth1980} Furthermore, the large measured magnetization suggests that domain wall pinning -- resulting from polycrystalline grains -- plays an important role, as it serves to increase the remanent magnetization above that of a uniform film (Fig.\,2c and Methods).

Beyond using an ac current source for magnetic actuation, we can also use the measured mechanical signal to amplify or dampen the resonator motion.  This type of feedback has been used in the cantilever sensing community, where amplification allows cantilever motion to be detected in highly damped conditions such as liquid environments\cite{Lar04} and cooling allows for faster measurements to be performed, due to the lower mechanical quality factor of the resonator, $Q_\textrm{m}$.  It is important to note that feedback cooling (or amplification) cannot increase the torsional sensitivity of the resonator, as the thermally-limited torque sensitivity on resonance is given by $S_\tau= 4k_{\rm B} T\Gamma I_{\rm eff}$, where $T$ is the mechanical mode temperature and $\Gamma = \Omega_{\rm m} / Q_{\rm m}$ is the mechanical damping rate.\cite{Kim16}  Feedback cooling lowers $T$ at the same rate that it increases $\Gamma$, therefore the torque sensitivity is unchanged.  Nonetheless, the change in the effective $Q$ can make a substantial difference in the visibility of the mechanical signal and the ring-up time of the mechanical mode.  It is the decrease in the ring-up time that we find particularly appealing for torque magnetometry of condensed matter systems.  For example, if one is interested in collective spin ensembles, in nano-magnetic\cite{Los15} or mesoscale superconducting\cite{Kim16} test samples, a lower $Q$ resonator -- while maintaining an equivalent torque sensitivity -- allows for measurement of higher frequency dynamics of spin modes.\cite{Los15}

\begin{figure*}[t]
\centerline{\includegraphics[width=5.44in]{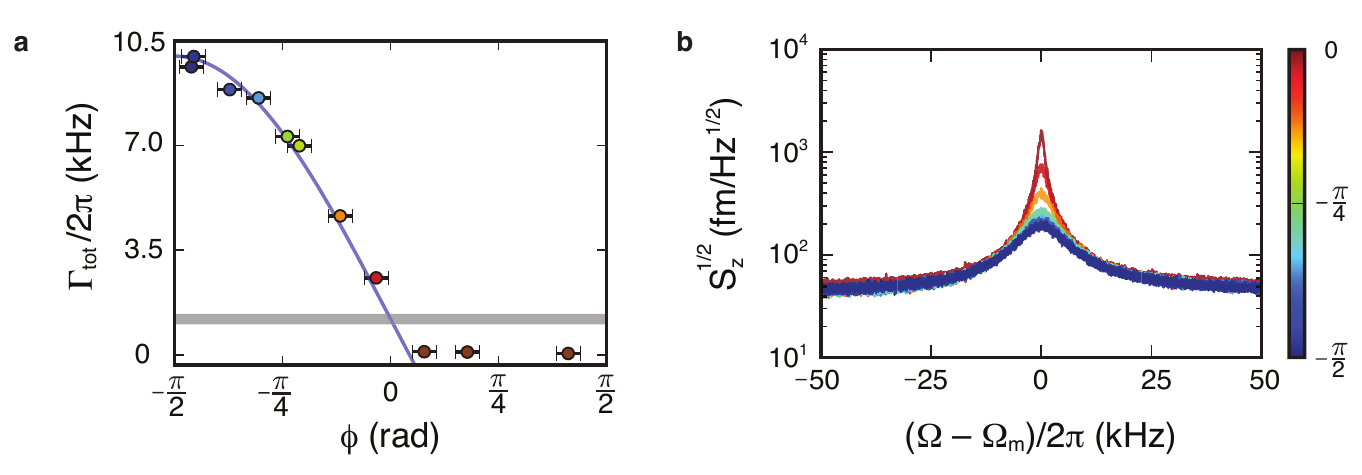}}
\caption{{\label{fig3}} \textbf{Magnetic feedback cooling and self-oscillations.} \textbf{a}, The total dissipation, $\Gamma_\textrm{tot}$, as a function of phase in the feedback loop, shows excellent agreement with a fit to Eqn.\,\ref{gamma} (blue curve). The optical power is 4.1 $\mu$W at the device. Maximum damping occurs at $-\pi/2$, and self-oscillation\cite{Kippenberg05} occurs for measured phases greater than zero (brown). The grey horizontal line shows the intrinsic dissipation, identical to the total dissipation at $\phi=0$. \textbf{b}, The same colour-coded data represented in terms of calibrated displacement spectral densities, for $\phi<0$.}
\end{figure*}

To test the performance of feedback amplification and cooling in our magneto-optomechanical system, we implement the scheme shown in Fig.\,1e.  Half of the ac signal, measuring the thermomechanical motion, is sent to a high-frequency data acquisition system, while the other half is phase shifted, bandpass filtered around the mechanical resonance of interest, and amplified, before being sent to the $z$-axis drive coil.  The phase shift is particularly important, as can be seen from the equation of motion for a damped, driven harmonic oscillator subject to a feedback force:

\begin{equation}
\ddot z + \Gamma_{\rm i} \dot z + \Omega_{\rm m}^2 z = \frac{F_\textrm{th}}{m_\textrm{eff}} - g_\textrm{fb}\Gamma_{\rm i} e^{i(\phi+{\pi}/{2})}(\dot z(t)+\dot z_{\rm n}(t)),
\label{fbeom}
\end{equation}

\noindent
where $\Gamma_\textrm{i}$ is the intrinsic mechanical dissipation rate.  The first term on the right hand side is proportional to the thermal force, $F_\textrm{th}$, acting on the mechanical resonator, while the second is related to the applied feedback force. Here $g_\textrm{fb}$ is the gain of the feedback loop, $\phi$ is the measured phase difference between the drive and the displacement, and $z_{\rm n}(t)$ is the measurement noise.\cite{Poggio07,Krause2015}  The resulting total mechanical dissipation can therefore be written as

\begin{equation}
\Gamma_{\textrm{tot}} = \Gamma_\textrm{i}(1+g_{\rm fb}\cos(\phi+\pi/2)).
\label{gamma}
\end{equation}

The phase controls whether the feedback gain results in cooling ($\phi=-\pi/2$), amplification ($\phi=\pi/2$), or has no effect ($\phi=0$).  In Fig.\,3, we show how the measured dissipation, $\Gamma_{\textrm{tot}}$, is affected by the phase in the feedback loop.  These measurements were performed with a moderate optical power of 4.1 $\mu$W at the microdisk, resulting in a feedback gain of $g_{\rm fb} = 7.4 \pm 0.1$, extracted from fitting Eqn.\,\ref{gamma} to the $\phi < 0$ data in Fig.\,3a.  We note that the feedback gain was kept moderate in these measurements to prevent overloading of the electronic components in the region of feedback amplification,  $\phi > 0$ data in Fig.\,3a, where the mechanical linewidth narrows and results in induced self-oscillation.\cite{Kippenberg05} 

\begin{figure}[t]
\centerline{\includegraphics[width=6in]{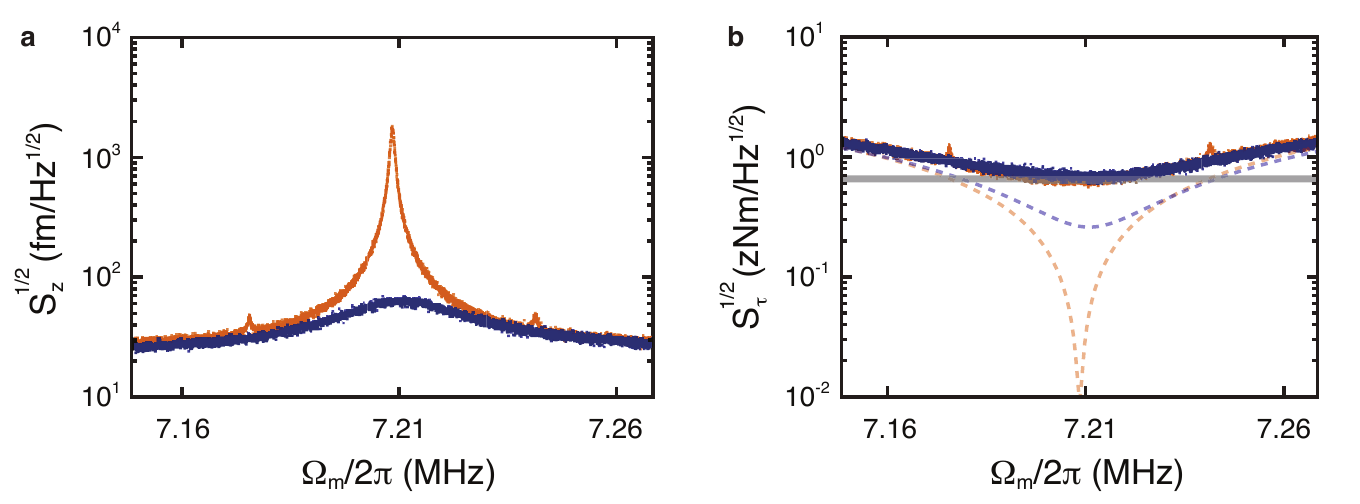}}
\caption{{\label{fig4}} \textbf{Optimal feedback cooling and torque sensitivity.} \textbf{a}, Increasing the power in the optical cavity enables sensitive mechanical transduction, lowering the imprecision noise floor to 25 fm/Hz$^{1/2}$ (orange trace).  The theoretical minimum temperature achievable with this noise floor is 8.6 K, and with a feedback phase of $-\pi/2$ and a feedback gain of 29 we find we are able to cool to 11.6 K (blue trace).  \textbf{b}, The calibrated torque sensitivity before (orange) and after (blue) feedback damping to a $Q_{\rm m}$ of just 260.  Dashed traces correspond to torque sensitivities in the absence of thermal noise, which is given by the grey line.  The device maintains a minimum torque sensitivity of 0.58 zNm/$\sqrt{\textrm{Hz}}$, regardless of whether or not feedback is applied.  This demonstrates the potential advantage of torque magnetometry with this system, which can maintain state-of-the-art torque sensitivity with short mechanical ring-up times.  }
\end{figure}

To demonstrate our full feedback cooling efficacy, we increase the optical power at the microdisk to 18 $\mu$W, which results in a shot-noise-limited noise floor of $S_z^{\rm imp}$ = 25 fm/Hz$^{1/2}$.  This optical power produces our largest feedback gain, as shown in Fig.\,4, increasing the total dissipation from $\Gamma_\textrm{i}/2\pi =930$ Hz ($Q_\textrm{m}$ = 7,760) to $\Gamma_\textrm{tot}/2\pi = 28,000$ Hz ($Q_\textrm{m}$ = 260). Evaluating Eqn.\,\ref{gamma} at $\phi = \pi/2$, a quantitative value for the feedback gain can be determined as $g_{\rm fb} = \Gamma_{\rm tot}/ \Gamma_{\rm i} - 1 = 29$. Furthermore, we can infer the reduced mechanical mode temperature using the relation\cite{Poggio07,Wilson2015}

\begin{equation}
T_\textrm{fb} = \left( T + \frac{m_{\rm eff} \Omega_{\rm m}^2 \Gamma_{\rm i} g_{\rm fb}^2 S_z^{\rm imp}}{4 k_{\rm B}} \right)\frac{1}{1+g_{\rm fb}},
\label{Tmeas}
\end{equation}

\noindent
for which we find we can cool from room temperature to a value of $T_\textrm{fb}$ = 11.6 $\pm$ 0.1 K. Note that the second term in Eqn.\,\ref{Tmeas} arises due to the fact that feeding back imprecision noise onto the resonator will act to heat the mechanical mode.  The  reduced mode temperature calculated from Eqn.\,\ref{Tmeas} is in reasonable agreement with that determined from the relative areas under the power spectral densities, which gives a mode temperature of 13.17 $\pm$ 0.02 K.  

Comparing with the theoretical minimum achievable temperature using feedback cooling, which for the high-temperature case ($k_{\rm B} T \gg \hbar \Omega_{\rm m}$) is given by

\begin{equation}
T_\textrm{min} = \sqrt{\frac{m_{\rm eff} \Omega_{\rm m}^2 \Gamma_{\rm i} T S_z^{\rm imp}}{k_{\rm B}}},
\label{Tmin}
\end{equation}

\noindent
we find $T_{\rm min}$ = 8.6 K for the values reported here. Note that this minimum temperature would be reached at a feedback gain of

\begin{equation}
g_{\rm fb,opt} = \sqrt{\frac{4 k_{\rm B} T}{m_{\rm eff} \Omega_{\rm m}^2 \Gamma_{\rm i} S_z^{\rm imp}}},
\label{gfbopt}
\end{equation}

\noindent
providing the optimal balance between cooling the mechanical motion via the cold damping feedback force and heating by feeding noise back onto the mechanical resonator. We calculate  $g_{\rm fb,opt}$ = 68 using the experimental parameters from the data in Fig.\,\ref{fig4}, approximately a factor of two above the measured value.  The result is a reduced temperature 3 K above the fundamental limit for the experiment performed here.

It is noteworthy that, as stated earlier, the torque sensitivity is not improved (or substantially degraded) by feedback cooling.  We show in Fig.\,4b the calibrated torque sensitivity using the room temperature data before cooling, and after cooling (damping) the torsional mode to 11.6 K.  Despite dramatically different dissipation rates and mechanical mode temperatures, they have identical torque sensitivities, with a minimum value of 0.58 zNm/Hz$^{1/2}$ -- a factor of two better than the only other room temperature cavity optomechanical torque sensor with an integrated nano-magnetic sample.\cite{Wu17}  Note that while cavity optomechanical resonators with integrated magnetostrictive materials, such as the one in Ref.~\citenum{Yu2016}, have achieved significantly better magnetic field sensitivity, they are not torque sensors and have no comparable torque sensitivity.

In summary, we have successfully integrated a ferromagnetic sample with a cavity optomechanical torsional resonator.  First, we used this magneto-optomechanical system as a magnetic field sensor with a linear response from 150 nT to 31 $\mu$T, and a responsivity of $0.134\pm0.003$ rad/mT.  Next, we quantitatively determined the magnetic properties of the magnetic sample, showing the potential of this system for studying mesoscopic condensed matter samples.  And finally, we showed that using magnetic actuation, the resonator motion could be driven into self-oscillation or cold-damped to below 12 K.  Future experiments could extend measurements to high frequency using torque-mixing,\cite{Los15} low temperatures for superconducting samples and enhanced sensitivity,\cite{Kim16} or explore quantum spin tunnelling\cite{Garanin2011} and exotic magnetic excitations such as those in topological systems.\cite{Qi09,Nagaosa13}

\section{Methods}
\subsection{Device Fabrication}
Starting with a silicon-on-insulator (SOI) chip, having a single-crystal silicon device layer of thickness 250 nm, fabrication requires two e-beam lithography (EBL) steps. First, EBL is used to pattern the silicon device shown in Fig.\,1a using ZEP-520a positive resist on a 30 kV EBL system (RAITH-150TWO). This is followed by a reactive ion etch (C$_4$F$_8$ and SF$_6$) to transfer the pattern to the silicon layer.  Afterwards, a second EBL process, with careful alignment, is used to pattern a PMMA bilayer resist in order to define the area for subsequent metal deposition. Electron-beam bombardment was used to deposit an 83 nm thick trilayer of Cr, Fe, and Cr. The purpose of the first 8 nm thick Cr layer is for adhesion to the silicon, whereas the last Cr layer, of equal thickness, serves as a capping layer to protect the iron from oxidation. Deposition took place at a pressure of $1.2\times 10^{-7}$ torr and a deposition rate of 0.9 \AA/s at 19 \degree C to minimize oxidation.  During deposition, a rare-earth magnet was placed directly underneath the 5 mm $\times$ 10 mm SOI chip to preferentially orient the iron magnetization direction along its length ($x$-axis in Fig.\,1a).  The magnetic field strength at the position of the device during deposition was measured to be 94 kA/m.

After lift-off of the PMMA and deposited metal using N-Methyl-2-pyrrolidone (NMP), the chip is placed in a vapour HF system (MEMStar Vapor HF) to etch the sacrificial SiO$_2$ layer.  Use of the HF vapour etch, instead of a buffered oxide etch, simultaneously avoids both problems of stiction and etching of the iron sample. It is also worth mentioning that the metallic needle cannot be arbitrarily long, as the stress induced during deposition causes deformation of the silicon resonator, apparent in Fig.\,1b. 

\subsection{Optomechanical Measurement}

System characterization is performed using a tuned-to-slope optomechanical detection scheme. \cite{Kim13,Kim16}  That is, as the torsional resonator moves in the evanescent field surrounding the WGM microdisk, the wavelength of the optical resonances are shifted.  This encodes information about the mechanical motion in the optical transmission through the cavity, measured using laser light from a dimpled tapered fibre.\cite{Hau14} The dimpled-tapered fibre has a radius of curvature of 70 $\mu$m, and touches the optical microdisk for stability and to ensure over-coupling ($\kappa_\textrm{e}/\kappa = 0.8$). The device presented here has an optomechanical coupling coefficient of $g_0=(d\omega_\textrm{c}/dz) z_\textrm{zpf}=$ 38 kHz, where $\omega_\textrm{c}$ is the optical cavity resonance frequency and the mechanical zero-point fluctuations are given by $z_\textrm{zpf}= \sqrt{\hbar/2 m_{\rm eff} \Omega_{\rm m}}=51$ fm.  The reasonably large $g_0$ enables measurement of the mechanical motion down to an imprecision noise-floor of 25 fm/Hz$^{1/2}$. During measurement, the device chip is mounted on a PCB with two-axes of orthogonal magnetic drive coils, Fig.\,1, although in the present experiment we only apply magnetic excitation along the $z$-axis.  We use a high-frequency lock-in amplifier to drive current through the $z$-axis excitation coil.  Measurements are performed in a room temperature optical-access vacuum chamber at $1\times10^{-5}$ torr. \cite{Hau14}  

For the data presented in Fig.\,3, the phase was varied by adding calibrated lengths of coaxial cable to the  feedback loop, and was measured using the lock-in amplifier at a frequency just below the mechanical resonance frequency.

\subsection{Magnetic Drive Calibration}
A Tektronix CT-2 current probe was used to calibrate the current output of the Zurich lock-in amplifier, which is then converted into a magnetic field ($\bm{H}_z$) at the position of the torsional device with the aid of a finite element simulation of the magnetic field, shown in Fig.\,1c. 

\subsection{Magnetic Simulation}
A GPU-accelerated open-source program, mumax (version 3.9), was used to simulate the zero temperature properties of a ferromagnetic needle with the as-fabricated dimensions of $4390 \times 410 \times 67$ nm$^3$ and the bulk properties of iron: exchange stiffness constant $A_{\rm ex} =$ 13.3 pJ/m and the $T=0$ saturation magnetization $M_{\rm s} = 1740$ kA/m,\cite{Wohlfarth1980} with a cell size of $10 \times 10 \times 6.7$ nm$^3$.  While the simulation does not take into account the effects of temperature, hence the $T=0$ value of the saturation magnetization is used, at room temperature the saturation magnetization is expected to be reduced by 2\% to 1710 kA/m.  Magnetocrystalline anisotropy is neglected on account of the large aspect ratio, which makes shape anisotropy dominant.\cite{Takahashi1962}  

The hysteresis loop starts at a large positive field, 80 kA/m, and the field is swept to -80 kA/m, finally returning to the original positive field.  A portion of the hysteresis loop, from -40 to 40 kA/m is presented in Fig.\,2c.  Hysteresis loops were simulated for both a uniform film, and one with a polycrystalline grain size of 40 nm, with a random variation of $M_s$ by $\pm10$\%.  The insets to Fig.\,2c show the corresponding magnetization states at $\pm$ 20 kA/m for a granulated structure.  The pristine structure shows additional domain states that are not included in Fig.\,2c.

\section{Acknowledgements}
This work was supported by the University of Alberta, Faculty of Science; the Natural Sciences and Engineering Research Council, Canada (Grants Nos. RGPIN-2016-04523, DAS492947-2016, and STPGP 493807 - 16); and the Canada Foundation for Innovation.  B.D.H. acknowledges support from the Killam Trusts.



\begin{thebibliography}{xxx}

\bibitem{Teu11}
J. D. Teufel, T. Donner, D. Li, J. W. Harlow, M. S. Allman, K. Cicak, A. J. Sirois, J. D. Whittaker, K. W. Lehnert and R. W. Simmonds. Sideband cooling of micromechanical motion to the quantum ground state. \textit{Nature} \textbf{475}, 359-363 (2011). 

\bibitem{Cha11}
J. Chan, T. P. Mayer Alegre, A. H. Safavi-Naeini, J. T. Hill, A. Krause, S. Gr\"{o}blacher, M. Aspelmeyer and O. Painter. Laser cooling of a nanomechanical oscillator into its quantum ground state. \textit{Nature} \textbf{478}, 89-92 (2011).

\bibitem{Oco10}
 A. D. O'Connell, M. Hofheinz, M. Ansmann, R. C. Bialczak, M. Lenander, E. Lucero, M. Neeley, D. Sank, H. Wang, M. Weides, J. Wenner, J. M. Martinis and A. N. Cleland. Quantum ground state and single-phonon control of a mechanical resonator. \textit{Nature} \textbf{464}, 697-703 (2010).

\bibitem{Cam11}
S. Camerer, M. Korppi, A. J\"{o}ckel, D. Hunger, T. W. H\"{a}nsch and P. Treutlein. Realization of an optomechanical interface between ultracold atoms and a membrane. \textit{Phys. Rev. Lett.} \textbf{107}, 223001 (2011).

\bibitem{Har16}
G. I. Harris, D. L. McAuslan, E. Sheridan, Y. Sachkou, C. Baser and W. P. Bowen. Laser cooling and control of excitations in superfluid helium. \textit{Nature Phys.} \textbf{12}, 788-793 (2016).

\bibitem{Forstner12}
S. Forstner, S. Prams, J. Knittel, E. D. van Ooijen, J. D. Swaim, G. I. Harris, A. Szorkovszky, W. P. Bowen, and H. Rubinsztein-Dunlop. Cavity optomechanical magnetometer. \textit{Phys. Rev. Lett.} \textbf{108}, 120801 (2012).

\bibitem{Wu17}
M. Wu, N. L.-Y. Wu, T. Firdous, F. Fani Sani, J. E. Losby, M. R. Freeman and P. E. Barclay. Nanocavity optomechanical torque magnetometry and radiofrequency susceptometry. \textit{Nature Nano.} \textbf{12}, 127-131 (2017).

\bibitem{Rath13}
P. Rath, S. Khasminskaya, C. Nebel, C. Wild and W. H. P. Pernice. Diamond-integrated optomechanical circuits. \textit{Nature Comm.} \textbf{4}, 1690 (2013).

\bibitem{Hoang16}
T. M. Hoang, Y. Ma, J. Ahn, J. Bang, F. Robicheaux, Z.-Q Yin and T. Li. Torsional Optomechanics of a Levitated Nonspherical Nanoparticle. \textit{Phys. Rev. Lett.} \textbf{117}, 123604 (2016).

\bibitem{Mitchell16}
M. Mitchell, B. Khanaliloo, D. P. Lake, T. Masuda, J. P. Hadden, P. E. Barclay. Single crystal diamond low-dissipation cavity optomechanics. \textit{Optica} \textbf{3}, 963-970 (2016).

\bibitem{Burek16}
M. J. Burek, J. D. Cohen, S. M. Meenehan, N. El-Sawah, C. Chia, T. Ruelle, S. Meesala, J. Rochman, H. A. Atikian, M. Markham, D. J. Twitchen, M. D. Lukin, O. Painter, and M. Lon\v{c}ar. Diamond optomechanical crystals. \textit{Optica} \textbf{3}, 1404-1411 (2016).

\bibitem{Li16}
T. Li and Z.-Q Yin. Quantum superposition, entanglement, and state teleportation of a microorganism on an electromechanical oscillator. \textit{Science Bulletin} \textbf{61}, 163 (2016).

\bibitem{Hil12}
J. T. Hill, A. H. Safavi-Naeini, J. Chan and O. Painter. Coherent optical wavelength conversion via cavity-optomechanics. \textit{Nature Comm.} \textbf{3}, 1196 (2012).

\bibitem{Los15}
J. E. Losby, F. Fani Sani, D. T. Grandmont, Z. Diao, M. Belov, J. A. J. Burgess, S. R. Compton, W. K. Hiebert, D. Vick, K. Mohammad, E. Salimi, G. E. Bridges, D. J. Thomson and M. R. Freeman. Torque-mixing magnetic resonance spectroscopy. \textit{Science} \textbf{350}, 798-801 (2015).

\bibitem{Bochmann13}
J. Bochmann, A. Vainsencher, D. D. Awschalom, and A. N. Cleland. Nanomechanical coupling between microwave and optical photons. \textit{Nature Phys.} \textbf{9}, 712-716 (2013).

\bibitem{Kim16}
P. H. Kim, B. D. Hauer, C. Doolin, F. Souris and J. P. Davis. Approaching the standard quantum limit of mechanical torque sensing. \textit{Nature Comm.} \textbf{7}, 13165 (2016).

\bibitem{Hauer13}
B. D. Hauer, C. Doolin, K. S. D. Beach and J. P. Davis. A general procedure for thermomechanical calibration of nano/micro-mechanical resonators \textit{Annals of Physics} \textbf{339}, 181-207 (2013).

\bibitem{Wohlfarth1980}
E. P. Wohlfarth. \textit{Ferro-magnetic Materials: A Handbook on the Properties of Magnetically Ordered Substances} (North Holland, Amsterdam, 1980).

\bibitem{Lar04}
N. V. Lavrik, M. J. Sepaniak, and P. G. Datskos. Cantilever transducers as a platform for chemical and biological sensors. \textit{Rev. Sci. Instrum.} \textbf{75}, 2229 (2004).

\bibitem{Poggio07}
M. Poggio, C. L. Degen, H. J. Mamin and D. Rugar. Feedback cooling of a cantilever's fundamental mode below 5 mK. \textit{Phys. Rev. Lett.} \textbf{99}, 017201 (2007).

\bibitem{Krause2015}
A. G. Krause, T. D. Blasius and O. Painter. Optical read out and feedback cooling of a nanostring optomechanical cavity. Preprint at \href{https://arxiv.org/pdf/1506.01249.pdf} {http://arxiv.org/abs/1506.01249} (2015).

\bibitem{Kippenberg05}
T. J. Kippenberg, H. Rokhsari, T. Carmon, A. Scherer and K. J. Vahala. Analysis of radiation-pressure induced mechanical oscillation of an optical microcavity. \textit{Phys. Rev. Lett.} \textbf{95}, 033901 (2005).

\bibitem{Wilson2015} 
D. J. Wilson, V. Sudhir, N. Piro, R. Schilling, A. Ghadimi and T. J. Kippenberg. Measurement-based control of a mechanical oscillator at its thermal decoherence rate. \textit{Nature} \textbf{524}, 325-329 (2015).

\bibitem{Yu2016}
C. Yu, J. Janousek, E. Sheridan, D. L. McAuslan, H. Rubinsztein-Dunlop, P. K. Lam, Y. Zhang and W. P. Bowen. Optomechanical Magnetometry with a Macroscopic Resonator. \textit{Phys. Rev. Applied} \textbf{5}, 044007 (2016).

\bibitem{Garanin2011}
D. A. Garanin and E. M. Chudnovsky. Quantum Entanglement of a Tunneling Spin with Mechanical Modes of a Torsional Resonator. \textit{Phys. Rev. X} \textbf{1}, 011005 (2011).

\bibitem{Qi09}
X.-L. Qi, R. Li, J. Zang and S.-C Zhang. Inducing a Magnetic Monopole with Topological Surface States. \textit{Science} \textbf{323}, 1184-1187 (2009).

\bibitem{Nagaosa13}
N. Nagaosa and Y. Tokura. Topological properties and dynamics of magnetic skyrmions. \textit{Nature Nano.} \textbf{8}, 899-911 (2013).

\bibitem{Kim13}
P. H. Kim, C. Doolin, B. D. Hauer, A. J. R. MacDonald, M. R. Freeman, P. E. Barclay and J. P. Davis. Nanoscale torsional optomechanics. \textit{Appl. Phys. Lett.} \textbf{102}, 053102 (2013).


\bibitem{Hau14}
B. D. Hauer, P. H. Kim, C. Doolin, A. J. R. MacDonald, H. Ramp and J. P. Davis. On-chip cavity optomechanical coupling. \textit{EPJ Tech. Instrum.} \textbf{1}:4 (2014).

\bibitem{Takahashi1962}
M. Takahashi. Induced Magnetic Anisotropy of Evaporated Films Formed in a Magnetic Field. \textit{Journal of Applied Physics} \textbf{33}, 1101-1106 (1962).


\end{thebibliography}
\end{document}